\documentclass[twocolumn,aps,showpacs,preprintnumbers,amsmath,amssymb, superscriptaddress]{revtex4-1}
\usepackage{bm}
\usepackage{graphicx}
\usepackage{color}
\usepackage{xspace}
\definecolor{dblue}{rgb}{0,0,0.6}
\definecolor{dred}{rgb}{0.9,0,0}
\definecolor{dgreen}{rgb}{0,0.4,0}

\pdfoutput=1

\begin{document}

\title{New phase transition in Na$_{2}$Ti$_{2}$As$_{2}$O revealed by Raman scattering}

\author{D.~Chen}
\affiliation{Beijing National Laboratory for Condensed Matter Physics, and Institute of Physics, Chinese Academy of Sciences, Beijing 100190, China}
\author{T.-T.~Zhang}
\affiliation{Beijing National Laboratory for Condensed Matter Physics, and Institute of Physics, Chinese Academy of Sciences, Beijing 100190, China}
\author{Z.-D.~Song}
\affiliation{Beijing National Laboratory for Condensed Matter Physics, and Institute of Physics, Chinese Academy of Sciences, Beijing 100190, China}
\author{H.~Li}
\affiliation{Beijing National Laboratory for Condensed Matter Physics, and Institute of Physics, Chinese Academy of Sciences, Beijing 100190, China}
\author{W.-L.~Zhang}
\affiliation{Beijing National Laboratory for Condensed Matter Physics, and Institute of Physics, Chinese Academy of Sciences, Beijing 100190, China}
\author{T.~Qian}
\affiliation{Beijing National Laboratory for Condensed Matter Physics, and Institute of Physics, Chinese Academy of Sciences, Beijing 100190, China}
\author{J.-L.~Luo}
\affiliation{Beijing National Laboratory for Condensed Matter Physics, and Institute of Physics, Chinese Academy of Sciences, Beijing 100190, China}
\affiliation{Collaborative Innovation Center of Quantum Matter, Beijing, China}
\author{Y.-G.~Shi}
\affiliation{Beijing National Laboratory for Condensed Matter Physics, and Institute of Physics, Chinese Academy of Sciences, Beijing 100190, China}
\author{Z.~Fang}
\affiliation{Beijing National Laboratory for Condensed Matter Physics, and Institute of Physics, Chinese Academy of Sciences, Beijing 100190, China}
\affiliation{Collaborative Innovation Center of Quantum Matter, Beijing, China}
\author{P.~Richard}\email{p.richard@iphy.ac.cn}
\affiliation{Beijing National Laboratory for Condensed Matter Physics, and Institute of Physics, Chinese Academy of Sciences, Beijing 100190, China}
\affiliation{Collaborative Innovation Center of Quantum Matter, Beijing, China}
\author{H.~Ding}\email{dingh@iphy.ac.cn}
\affiliation{Beijing National Laboratory for Condensed Matter Physics, and Institute of Physics, Chinese Academy of Sciences, Beijing 100190, China}
\affiliation{Collaborative Innovation Center of Quantum Matter, Beijing, China}

\date{\today}

\begin{abstract}
We performed a Raman scattering study of Na$_2$Ti$_2$As$_2$O. We identified a symmetry breaking structural transition at around $T_s = 150$ K, which matches a large bump in the electrical resistivity. Several new peaks are detected below that transition. Combined with first-principles calculations, our polarization-dependent measurements suggest a charge instability driven lattice distortion along one of the Ti-O bonds that breaks the 4-fold symmetry and more than doubles the unit cell.
\end{abstract}



\maketitle

Because their layered structure resembles both the copper oxide and the iron-based superconductors, special attention is devoted to the $A$Ti$_{2}Pn_{2}$O ($A$ = Na$_2$, Ba, $Pn$ = As, Sb, Bi) system \cite{review}, in particular regarding anomalies in their transport properties that may be caused by orders competing with superconductivity. Although the maximum $T_c$ reported among these compounds is only 6.1 K in Ba$_{1-x}$K$_x$Ti$_2$Sb$_2$O ($x$ = 0.12), their phase diagram points towards a competition between density-wave states and superconductivity \cite{Ba1-xKxTi2Sb2O6.1K,Ba1-xNaxTi2Sb2O5.5K}. However, the origin of the density-wave remains controversial. While some calculations attribute the anomalies of their transport properties to a spin-density-wave (SDW) transition \cite{calcualtion_SDW,BaAs_calcaulation_SDW,luzhongyi_SDW}, most experiments rather favor a charge-density-wave (CDW) scenario \cite{nonmagnetic_NMR,nonmagnetic_usr}. Moreover, nuclear quadruple resonance (NQR) experiments show that the in-plane 4-fold symmetry at the Sb site is broken below the transition temperature in BaTi$_2$Sb$_2$O \cite{NQR_symmetrybroken}. Recently, neutron diffraction experiments suggest an ``intra-unit-cell nematic charge order" in BaTi$_2$As$_2$O \cite{unitcell_nemeticity}, which makes the titanium oxypnictides more intriguing.

In this Rapid Communication we focus on Na$_2$Ti$_2$As$_2$O, which has the highest CDW transition temperature among this series \cite{first_synthesis}. Previous optical spectroscopy measurements indicate two CDW transitions at 42 K and 320 K in this material \cite{NaAs_IR}, while there is only little known about the lattice dynamics. Our Raman scattering study of Na$_2$Ti$_2$As$_2$O is supported by first-principles calculations of the vibration modes. We observe a total of 4 single-phonon modes predicted by our first-principles calculations in the high-temperature phase and 1 additional feature assigned to a double-phonon excitation. By doing temperature-dependent Raman measurements, we reveal a structural transition breaking the 4-fold symmetry at around $T_s$ = 150 K, which is consistent with a yet not understood anomaly in the electrical resistivity. Combining the results of our experiments with calculations, we suggest that this structural transition does not rotate the high-symmetry axes but more than doubles the primitive cell.

\begin{figure*}[!t]
\begin{center}
\includegraphics[width=\textwidth]{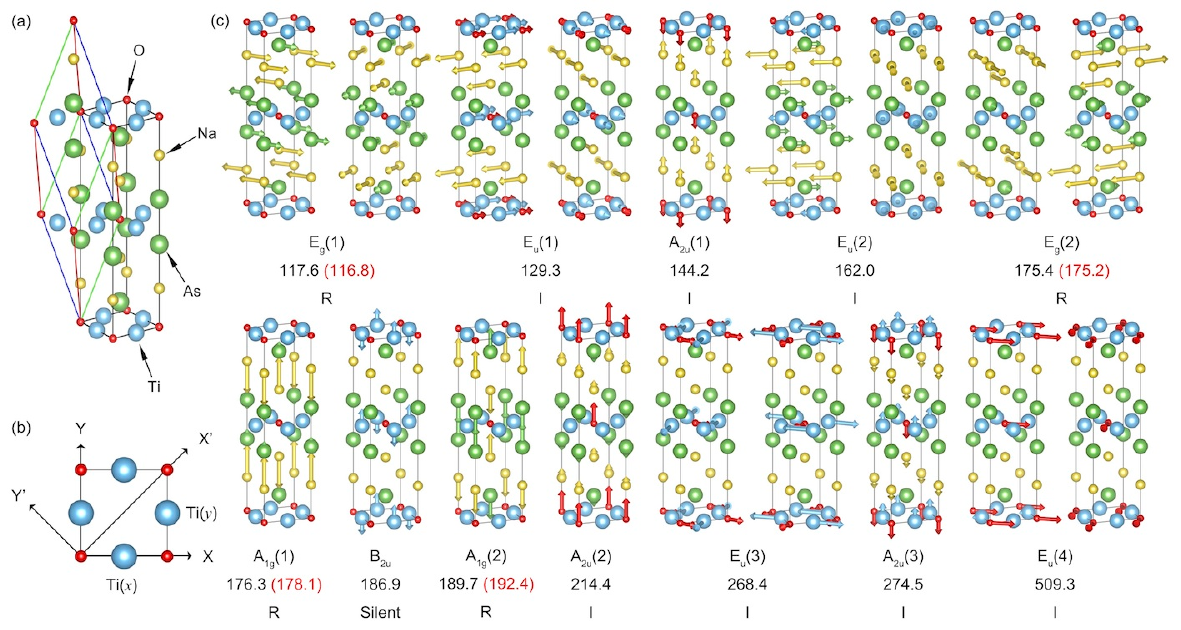}
\end{center}
\caption{\label{figure1}(Color online). (a) Crystal structure of Na$_{2}$Ti$_{2}$As$_{2}$O. The black rectangular prism shows the unit cell and the colored parallelepiped shows the primitive cell. (b) Ti-O plane and definitions of directions used in this paper. (c) Main atomic displacements for the optical modes of Na$_{2}$Ti$_{2}$As$_{2}$O. The displacements of the atoms are indicated by arrows. The first line below each configuration of vibration indicates the mode symmetry in the $D_{4h}$ group notation. Numbers in parenthesis are used to specify modes when there are more than one mode with the same symmetry. The second line below each configuration of vibration indicates the calculated (experimental) mode energy. The last line below each configuration of vibration gives the optical activity, with I = infrared active, R = Raman active, Silent = not optically active. }
\end{figure*}

The Na$_2$Ti$_2$As$_2$O single crystals used in our measurements were grown by the flux method \cite{shiyouguo}. Freshly grown plate-like samples with typical size of $1\times2\times0.06$~mm$^{3}$ were prepared and electrodes were glued on them in a glove box for avoiding air contamination. The in-plane resistivity was measured using a commercial physical properties measurement system (PPMS). Raman scattering measurements were performed using 488.0 nm and 514.5 nm laser excitations in a back-scattering micro-Raman configuration with a triple-grating spectrometer (Horiba Jobin Yvon T64000) equipped with a nitrogen-cooled CCD camera. The crystals were cleaved by tape to obtain clean and flat surfaces, and then transferred into a low-temperature cryostat for Raman measurements between 10 and 350 K with a working vacuum better than $2\times 10^{-6}$ mbar. A $50\times$ long-focus distance objective was used to both focus the laser beam and collect the scattered light. The power at the sample was kept small and the laser heating was negligible according to our tests. As shown in Fig. \ref{figure1}(b), we define X and Y as the directions along the Ti-O bonds, and X' and Y' as the directions along the Ti-Ti bonds, which form a 45$^\circ$ angle with the Ti-O bonds. The Z direction corresponds to the axis perpendicular to the Ti$_2$O planes at room temperature.

\begin{figure*}[!t]
\begin{center}
\includegraphics[width=\textwidth]{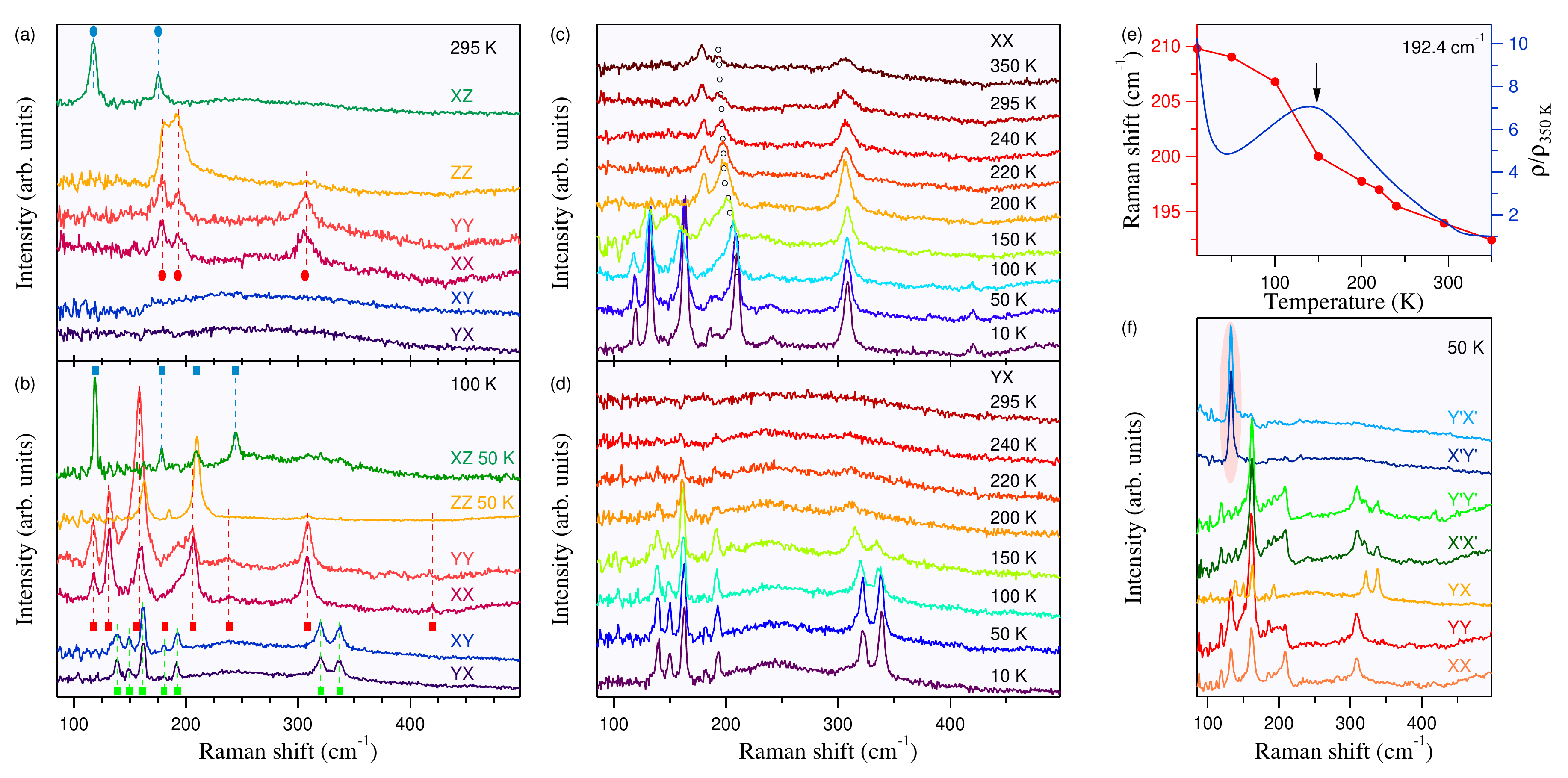}
\end{center}
\caption{\label{figure2}(Color online). (a) Raman spectra under different incident and scattered polarizations at room temperature. The blue circles indicate E$_{g}$ modes, whereas the red circles indicate A$_{1g}$ modes. (b) Same as (a) but at 100 K (except that the spectra under (XZ) and (ZZ) polarizations are measured at 50 K). The squares with different colors indicate modes with different symmetries after the phase transition. The curves are shifted relative to each other for clarity. (c) Waterfall plot of spectra recorded under the (XX) polarization configuration across the temperature range studied. The circles indicate the evolution of the A$_{1g}(2)$ peak at 192.4~cm$^{-1}$. (d) Same as (c) but under the (YX) polarization configuration. (e) In-plane resistivity of Na$_{2}$Ti$_{2}$As$_{2}$O normalized by the resistivity at 350 K (in blue), and temperature dependence of the A$_{1g}(2)$ peak position (in red). The arrow indicates the anomaly discussed in the text. (f) Raman spectra under different incident and scattered polarizations at 50 K. The pink shadow indicates the A$_g$ mode at 132.0~cm$^{-1}$. }
\end{figure*}

The Na$_{2}$Ti$_{2}$As$_{2}$O crystal structure is characterized by space group $I4/mmm$ (point group $D_{4h}$) at room temperature \cite{shiyouguo}. A simple group symmetry analysis \cite{bilbal} indicates that the phonon modes at the Brillouin zone (BZ) center $\Gamma$ decompose into [2A$_{1g}$+2E$_{g}$]+[3A$_{2u}$+B$_{2u}$+ 4E$_{u}$]+[A$_{2u}$+E$_{u}$], where the first, second and third terms represent the Raman-active modes, the infrared-active modes and the acoustic modes, respectively. To get estimates on the mode frequencies of Na$_{2}$Ti$_{2}$As$_{2}$O, we employed VASP and set a $6\times6\times6$ Monkhorst-pack $k$ point mesh and a 400 eV cutoff for wavefunctions. Using the generalized gradiant approximation \cite{GGA}, we relaxed both the cell and atom coordinates of experimental results \cite{first_synthesis} until the change in the total energy becomes smaller than $10^{-6}$ eV. Knowing the information on the ground state, it is easy to use the Phonopy package, which implements the density functional perturbation theory \cite{DFPT1,DFPT2,theory2}, to get the phonon frequencies and vibration modes at the $\Gamma$ point. All the optic vibration modes, along with the corresponding irreducible representations, the comparison between calculated and experimental phonon frequencies, as well as the optical activity of the modes, are given in Fig. \ref{figure1}(c).

\begin{figure}[!t]
\begin{center}
\includegraphics[width=\columnwidth]{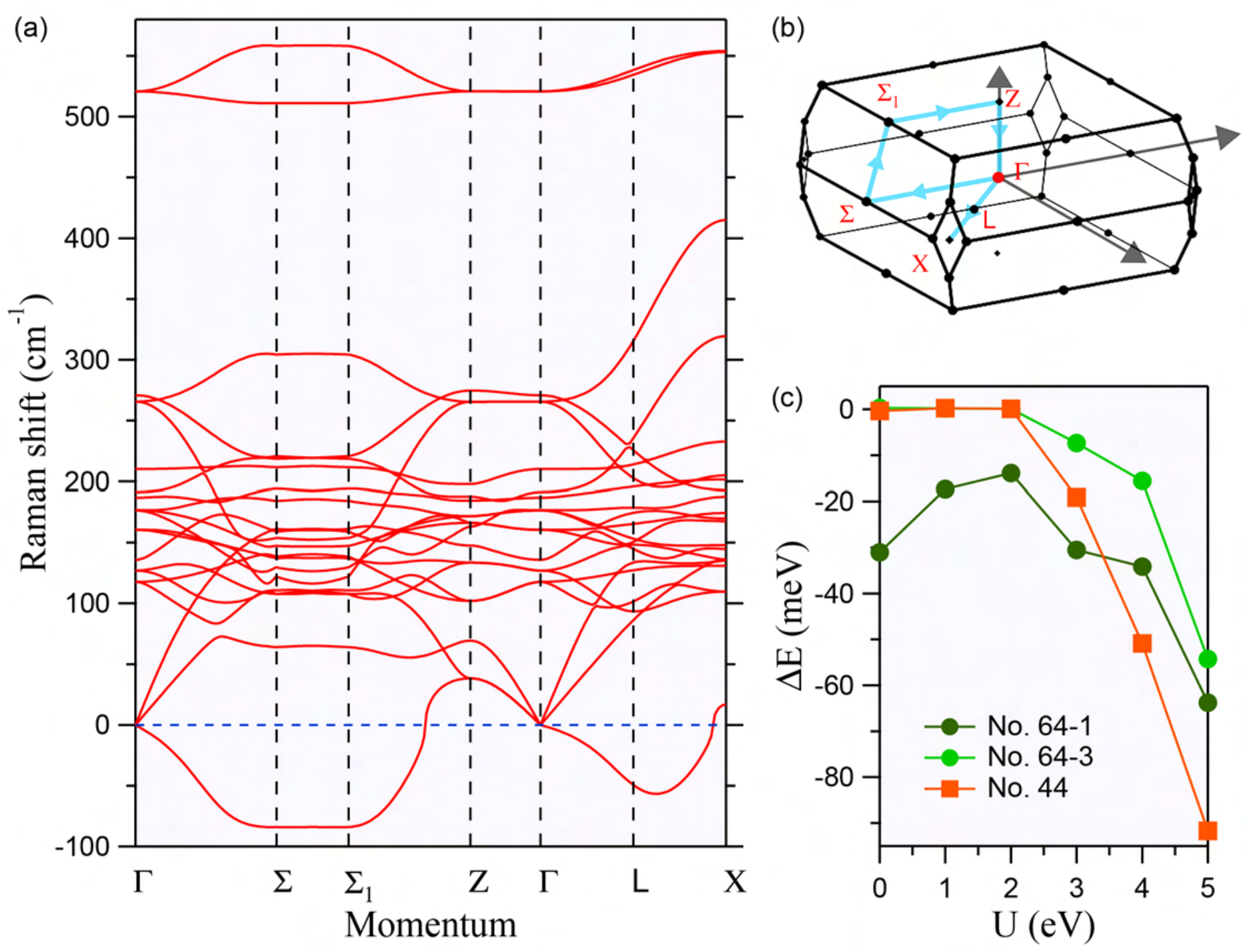}
\end{center}
\caption{\label{figure3}(Color online). (a) Calculation of the phonon band dispersions in the normal state of Na$_{2}$Ti$_{2}$As$_{2}$O along high-symmetry lines. The imaginary frequencies are shown as negative. (b) Definition of high-symmetry points in the momentum space. The blue arrows show the momentum path corresponding to the dispersion displayed in (a). (c) Calculation of the energy difference between space groups No. $64$, No. $44$ space group No. $139$ (normal structure) as a function of the Coulomb repulsion energy U. }
\end{figure}

The Raman tensors corresponding to the $D_{4h}$ symmetry group are expressed in the XYZ coordinates as (note that the B$_{1g}$ and B$_{2g}$ channels are absent in Na$_{2}$Ti$_{2}$As$_{2}$O):
\begin{displaymath}
\textrm{A$_{1g}=$}
\left(\begin{array}{ccc}
a & 0 &0\\
0 & a &0\\
0 & 0 &b
\end{array}\right)
, \left[\begin{array}{ccc}\textrm{E$_{g}=$}
\left(\begin{array}{ccc}
0 & 0 &e\\
0 & 0 &0\\
e & 0 &0\\
\end{array}\right)
, \left(\begin{array}{ccc}
0 & 0 &0\\
0 & 0 &e\\
0 & e &0\\
\end{array}\right)
\end{array}\right].
\end{displaymath}
\noindent Using the polarization selection rules and the fact that the Z axis is obviously recognized by the morphology of the samples, the assignments of the Raman symmetries for each mode is straightforward. In Fig. \ref{figure2}(a), we show the Raman spectra of Na$_{2}$Ti$_{2}$As$_{2}$O recorded at room temperature under the $(\mathbf{\hat{e}}^i\mathbf{\hat{e}}^s)=$ (XZ), (ZZ), (YY), (XX), (XY) and (YX) polarization configurations, where $\mathbf{\hat{e}}^i$ and $\mathbf{\hat{e}}^s$ are the incident and scattered light polarizations, respectively. We observe a total of 5 modes. As indicated in Fig. \ref{figure1}(c), the experimental mode frequencies are in very good agreement with the calculated modes, except for the mode at 306.7~cm$^{-1}$, which cannot be assigned unambiguously, but may be related to a double phonon excitations due to its broadness. For perfectly aligned crystals, pure A$_{1g}$ symmetry is obtained in the parallel polarization configurations, in which we detect 2 A$_{1g}$ peaks at 178.1~cm$^{-1}$ and 192.4~cm$^{-1}$. They correspond to vibrations of atoms along the Z axis. Pure E$_{g}$ symmetry is obtained in the (XZ) configuration. In this channel, we observe 2 peaks at 116.8~cm$^{-1}$ and 175.2~cm$^{-1}$, both corresponding to vibrations of atoms in the XY plane. No peak was detected in the (XY) and (YX) channels. As expected, the peak positions and intensities obtained in the (XX) and (YY) channels are exactly same, indicating the perfect 4-fold symmetry of Na$_{2}$Ti$_{2}$As$_{2}$O at room temperature.


As shown in Figs. \ref{figure2}(c) and \ref{figure2}(d), the Raman spectra vary significantly with temperature decreasing. New peaks begin to emerge at around 220 K under (XY) polarization. The A$_{1g}$ peaks detected at room temperature under (XX) polarization remain down to the lowest temperature, although their spectral intensity is strongly affected at around 150 K, temperature at which new peaks become visible. Besides peak intensities and linewidths, there is no further change upon cooling further below 150 K. As an example, we display in Fig. \ref{figure2}(e) the temperature evolution of the A$_{1g}(2)$ mode. A clear discontinuity occurs at 150 K. Although most debates on density-waves in Na$_{2}$Ti$_{2}$As$_{2}$O concern possible transitions at 42 K and 320 K, it is clear from our Raman data that a structural transition takes place at around $T_s=150$ K. Interestingly, this temperature coincides with a large bump in the resistivity data shown in Fig. \ref{figure2}(e).


In Fig. \ref{figure2}(b) we display the Raman spectra of Na$_{2}$Ti$_{2}$As$_{2}$O recorded below $T_s$. There are at least 4 peaks (119.1~cm$^{-1}$, 178.2~cm$^{-1}$, 209.8~cm$^{-1}$ and 244.3~cm$^{-1}$) in the (XZ) channel, 8 peaks (116.8~cm$^{-1}$, 132.0~cm$^{-1}$, 158.5~cm$^{-1}$, 181.2~cm$^{-1}$, 206.4~cm$^{-1}$, 238.3~cm$^{-1}$, 309.0~cm$^{-1}$ and 420.1~cm$^{-1}$) in the (XX), (YY) and (ZZ) channels, and 7 peaks (138.1~cm$^{-1}$, 149.5~cm$^{-1}$, 162.3~cm$^{-1}$, 180.4~cm$^{-1}$, 191.7~cm$^{-1}$, 320.1~cm$^{-1}$ and 337.9~cm$^{-1}$) in the (XY) and (YX) channels, for a total of 19 modes, which is more than the total number of optic modes at room temperature, implying that the primitive cell enlarged. Interestingly, the peak intensities for the (XX) and (YY) channels are no longer identical, indicating the breakdown of the 4-fold symmetry. Along with the lack of coincident peaks in the (XX), (XY) and (XZ) channels, the breakdown of the 4-fold symmetry restricts the possible point groups for the low-temperature phase of Na$_{2}$Ti$_{2}$As$_{2}$O to D$_{2h}$, C$_{2v}$ and D$_{2}$. In order to extract more information from the spectra, let's first assume a D$_{2h}$ symmetry. The Raman tensors associated with the D$_{2h}$ point group are:
\begin{displaymath}
\textrm{A$_{g}$=}
\left(\begin{array}{ccc}
a & 0 &0\\
0 & b &0\\
0 & 0 &c
\end{array}\right)
,\textrm{B$_{1g}$=}
\left(\begin{array}{ccc}
0 & d &0\\
d & 0 &0\\
0 & 0 &0
\end{array}\right),
\end{displaymath}
\begin{displaymath}
\textrm{B$_{2g}$=}
\left(\begin{array}{ccc}
0 & 0 &e\\
0 & 0 &0\\
e & 0 &0
\end{array}\right)
,\textrm{B$_{3g}$=}
\left(\begin{array}{ccc}
0 & 0 &0\\
0 & 0 &f\\
0 & f &0
\end{array}\right).
\end{displaymath}
\noindent As shown in Fig. \ref{figure2}(f), the (X'X') and (Y'Y') spectra are almost identical and the peaks showing up are a combination of peaks in the (XX) and (YY) channels, in agreement with the Raman selection rules for the D$_{2h}$ group presented in Table \ref{D2h}. Since these Raman selection rules are given by defining X and Y as the unit cell axes, the transition here does not change the direction of the high-symmetry axes. Consequently, the phase transition at $T_s$ in Na$_{2}$Ti$_{2}$As$_{2}$O enlarges the unit cell along one of the Ti-O directions, inducing a 2-fold symmetry.

\begin{table*}
\caption{\label{D2h}Raman selection rules for A$_{g}$ and B$_{1g}$ modes of $D_{2h}$ point group}
\begin{ruledtabular}
\begin{tabular}{cccccccccc}
 Modes& (XX)& (YY)& (XY)& (YX)& (X'X')& (Y'Y')& (X'Y')& (Y'X')\\
\hline
A$_{g}$& a$^2$& b$^2$& - & -& (a+b)$^2$/4& (a+b)$^2$/4& (a-b)$^2$/4& (a-b)$^2$/4\\
B$_{1g}$& -& -& d$^2$& d$^2$& d$^2$& d$^2$& -& -\\
\end{tabular}
\end{ruledtabular}
\end{table*}

According to the Raman selection rules shown in  Table \ref{D2h}, the intensities of the A$_g$ peaks should be proportional to (a-b)$^2$/4 under the (X'Y') and (Y'X') polarization configurations, in contrast to (a+b)$^2$/4 under the (X'X') and (Y'Y') polarization configurations. A comparison of the spectra in Fig. \ref{figure2}(f) shows that most of the A$_g$ peaks have similar intensities in the (X'X'), (Y'Y'), (XX) and (YY) channels, while they have nearly zero intensity in the (X'Y') and (Y'X') channels, indicating that $a\thickapprox b$. The A$_g$ peak at 132.0~cm$^{-1}$ has the inverse behavior, and thus $a\thickapprox -b$ for this peak. The conclusion of this little analysis is that the structural distortion bringing the system from a 4-fold symmetry to a 2-fold symmetry, is actually small, and it is not so surprising that it was not identified by previous XRD measurements \cite{XRD_As_Sb_135Ktransition}.

We now focus on the origin of this phase transition. In Fig. \ref{figure3}(a), we display our calculations of the phonon dispersion along high-symmetry lines for the normal state Na$_{2}$Ti$_{2}$As$_{2}$O. A conspicuous feature is the presence of imaginary modes, indicating an unstable lattice. According to the locations of the imaginary modes, the CDW transition ocuring at 320 K may have possible wave vectors $\mathbf{q}$ equal to $\Sigma$ ($\frac{1}{2}$, 0, 0), $\Sigma_1$ ($\frac{1}{2}$, 0, $\frac{1}{2}$) or $L$ ($\frac{1}{2}$, $\frac{1}{2}$, 0). Although it is hard to match the phonon energy of the emerging peaks because of the imaginary modes, the symmetry of the wave vectors ($\Sigma$ (D$_{2h}$), $\Sigma_1$ (C$_{2h}$) and $L$ (C$_{2v}$)) could drive the structural transition observed. We also found that the imaginary modes mainly come from unequal vibrations of Ti($x$) and Ti($y$) atoms. Since the electrons at the Fermi surface mainly derive from Ti $3d$ bands \cite{calculationofBaNaAsSb_3dorbital}, we conclude that the charge instability on Ti($x$) and Ti($y$), that possibly starts developing at 320 K, may drive the structural transition at 150 K. As shown in Fig. \ref{figure2}(e), the structural transition we observe explains the resistivity anomaly and corresponds well to the dramatic change at 150 K in the Hall coefficient \cite{Hall}.

Recently, a structural transition from $P4/mmm$ ($D_{4h}^1$) to $Pmmm$ ($D_{2h}^1$) at the same temperature as a resistivity anomaly was reported in BaTi$_{2}$As$_{2}$O \cite{unitcell_nemeticity}. Since no CDW superlattice peak was observed in this material, it was suggested that an intra-unit-cell nematic charge order model is driving the structural transition. Although what happens here in Na$_{2}$Ti$_{2}$As$_{2}$O is somehow similar since the 4-fold symmetry is lowered to a 2-fold symmetry without changing the unit cell axes directions, in Na$_{2}$Ti$_{2}$As$_{2}$O the crystal symmetry also changes and the primitive cell become larger. To find out what kind of structural transition corresponds to Na$_{2}$Ti$_{2}$As$_{2}$O, we analyzed its subgroups for a given $k$-index defining the multiplication of the cell, and a $t$-index defining the ratio between the order of the original group and the order of the subgroup \cite{subgroup}. 22 kinds of subgroups were obtained under the conditions $t$ = 1, 2 and $k$ = 1, 2, among which only the ground energies of space groups No. 44 (c, a, b), No. 64-1 (c, a+b, -a+b), No. 64-3 (-a-b, a-b, c) are lower than the normal state of Na$_{2}$Ti$_{2}$As$_{2}$O according to the first-principles pseudopotential plane wave method package in VASP \cite{VASP1,VASP2}. We present the energy difference between these 3 kinds of subgroups and the normal state of Na$_{2}$Ti$_{2}$As$_{2}$O (No. 139) considering different values of the Coulomb repulsion energy U of the Ti $3d$ electrons in Fig. \ref{figure3}(c). Unfortunately, space group No. 44 does not enlarge the primitive cell, and space groups No. 64-1 and No. 64-3 rotate the unit cell axes by 45$^\circ$. Therefore, none of potential space group candidates matches the experiments for $t$ = 1, 2 and $k$ = 1, 2. Consequently, we conclude that the structural transition at $T_s$ likely leads to a multiplication of the unit cell by more than a factor of 2.

It is worth noting that in most of the titanium oxypnictides the CDW transitions or structural transitions are difficult to detect experimentally. For example, only small indications pointing to a possible CDW transition have been obtained by Raman scattering on Ba$_2$Ti$_2$Fe$_2$As$_4$O \cite{shangfei}, the intergrowth of BaTi$_2$As$_2$O and BaFe$_2$As$_2$. Our study on Na$_{2}$Ti$_{2}$As$_{2}$O is a clear counter example that may serve as an ideal playground to study the interplay between charge order, lattice and superconductivity in the titanium oxypnictides.


We acknowledge S.-K. Su, Y.-H. Gu, P. Zhang, X. Shi for useful discussions. This work was supported by grants from MOST (2011CBA001000, 2011CBA00102, 2012CB821403, 2013CB921703 and 2015CB921301) and NSFC (11004232, 11034011/A0402, 11234014, 11274367, 11474330, 11274362 and 11474330) from China, as well as by the Strategic Priority Research Program (B) of the Chinese Academy of Sciences (No. XDB07020000).

\bibliography{citation}

\end{document}